\documentclass[aps,prc,twocolumn,superscriptaddress,showpacs,10pt,dvipsnames]{revtex4-1}

\usepackage[utf8]{inputenc}
\usepackage{amsfonts}
\usepackage{amsmath}
\usepackage{amssymb}
\usepackage{bm}
\usepackage[pdftex]{graphicx}
\usepackage{xcolor}
\usepackage[normalem]{ulem}
\usepackage{comment}
\usepackage{multirow}

\begin{document}


\title{Three-body structure of $^{19}$B: Finite-range effects in two-neutron halo nuclei}




\author{J. Casal}
\email{casal@pd.infn.it}
\affiliation{Dipartimento di Fisica e Astronomia ``G.Galilei'', Università degli Studi di Padova, via Marzolo 8, Padova, I-35131, Italy}
\affiliation{INFN-Sezione di Padova, via Marzolo 8, Padova, I-35131, Italy}
\author{E. Garrido}
\affiliation{Instituto de Estructura de la Materia, IEM-CSIC, Serrano 123, E-28006 Madrid, Spain}


\date{\today}

\begin{abstract}
The structure and $B(E1)$ transition strength of $^{19}$B are investigated in a $^{17}\text{B}+n+n$ model, triggered by a recent experiment showing that $^{19}$B exhibits a well pronounced two-neutron halo structure. 
Preliminary analysis of the experimental data was performed by employing contact $n$-$n$  interactions, which are known to underestimate the $s$-wave content in other halo nuclei such as $^{11}$Li. In the present work, the three-body hyperspherical formalism with finite-range two-body interactions is used to describe $^{19}$B. 
In particular, two different finite-range $n$-$n$ interactions will be used, as well as a simple central Gaussian potential whose range is progressively reduced. The purpose is to determine the main properties of the nucleus and investigate how they change when using contact-like $n$-$n$ potentials. Special attention is also paid to the dependence on the prescription used to account for three-body effects, i.e., a three-body force or a density-dependent $n$-$n$ potential.
We have found that the three-body model plus finite-range potentials provide a description of $^{19}$B consistent with the experimental data. The results are essentially independent of the short-distance details of the two-body potentials, giving rise to an $(s_{1/2})^2$ content of about 55\%, clearly larger than the initial estimates. Very little dependence has been found as well on the prescription used
for the three-body effects. 
The total computed $B(E1)$ strength is compatible with the experimental result, although we slightly overestimate the data around the low-energy peak of the $dB(E1)/d\varepsilon$ distribution. 
Finally, we show that a reduction of the $n$-$n$ interaction range produces a significant reduction of the $s$-wave contribution, which then should be expected in calculations using contact interactions.
\end{abstract}


\maketitle



The story of halo nuclei started back in the late 1980s, when an unexpectedly large interaction cross section for some particular light nuclei isotopes was observed \cite{Tanihata1985}. Among them, the cases of $^{11}$Li and $^6$He are the most prominent examples of the so-called two-neutron halo nuclei, where the two outer neutrons reside basically in the classically forbidden region, far apart from the core of the nucleus \cite{Riisager1994}. This fact intuitively suggests that these systems could be described as clusterized structures, where the internal degrees of freedom of the clusters are frozen. In fact, for the particular case of two-neutron halo nuclei, this kind of structure was soon suggested in Ref.~\cite{Hansen1987}, where $^{11}$Li was for the first time described as a bound three-body system that could be interpreted as a dineutron coupled to the nuclear core.

It is clear that, within this picture, the key ingredients are the two-body interactions between the different clusters, that is, for two-neutron halo systems, the core-neutron and the neutron-neutron potentials. Typically, the core-neutron interactions are taken as phenomenological potentials whose parameters are adjusted to reproduce the available experimental information on the core-neutron subsystem. A correct reproduction of the core-neutron energy spectrum, i.e., bound states, resonances, or virtual states, is crucial. Together with the two-body potentials, an additional required  ingredient comes from the fact that the use of bare two-body interactions is usually not enough to reproduce the experimental separation energy. This deficiency is commonly repaired by introducing an effective three-body force (see, e.g., Refs.~\cite{Zhukov93,IJThompson04}), which, in principle, takes care of all those effects, such as polarization of the clusters, that go beyond pure two-body correlations.

In any case, it is known that for well-extended systems, such as halo nuclei, the details of the potentials are not very relevant, and different potential shapes reproducing the same low-energy two-body properties provide quite similar results. This is shown for instance in Ref.~\cite{Garrido04}, where the structure of $^{17}$Ne is found to be very similar with a Gaussian or Woods-Saxon neutron-core potential, or even when using a homemade neutron-neutron interaction compared to the more sophisticated Argonne potential~\cite{av18}. The only common property of all these potentials is their finite-range character (except the Coulomb interaction in those cases where more than one charged cluster is involved).

This contrasts with the rather frequent use of contact interactions, which provide the nice feature of simplifying the calculations, since many of the integrations involved become analytical. In particular, the use of a zero-range neutron-neutron interaction makes easier the investigation of pairing phenomena and dineutron correlations \cite{Esbensen1997,Hagino05}. This interaction is often a density-dependent force, which, in practice, plays the role of the effective three-body force, and at the same time simulates the modification of the bare two-body interaction due to the presence of the core. 
In general, this kind of potentials permits to obtain a rather accurate picture of the structure and gross properties of halo nuclei. However, this could be not true anymore when going down to the details of the structure. This is pointed out 
in Ref.~\cite{Esbensen1997}, where the too low $s$-wave content in 
$^{11}$Li ($\approx 23$\%) compared to the experiment \cite{Young1994}
is discussed. Ulterior experiments and theoretical calculations \cite{Zinser1997,Garrido1997,Casal2017a,GomezRamos17a} confirmed this fact.
More recently, Oishi \textit{et al.}~\cite{Oishi2017} concluded that the use of a schematic density-dependent contact nucleon-nucleon potential makes not possible
simultaneous reproduction of the empirical $Q$ value, decay width, and the nucleon-nucleon scattering length in the two-proton decay of $^6$Be. A more sophisticated neutron-neutron interaction is needed.

In a recent paper, the structure of $^{19}$B has been experimentally investigated by means of exclusive measurements of its Coulomb dissociation, in collision on a lead target, into  $^{17}$B and two neutrons \cite{Cook2020}. The enhanced electric dipole strength observed constitutes a clear evidence of the presence of a prominent two-neutron halo in the system. The structure properties of $^{19}$B arising from the provided experimental information are obtained by using the contact density dependent neutron-neutron interaction employed in Refs.~\cite{Esbensen1997,Hagino05}.

The purpose of this work is to investigate $^{19}$B, understood as $^{17}\text{B}+n+n$, by use of finite-range two-body interactions, and analyze the consequences that a progressive reduction of the neutron-neutron potential range, eventually up to zero-range, has on the structure of the system. Also, special attention is paid to the role played by the
three-body force and the effect of replacing it by a density-dependent term in the neutron-neutron potential. 


\paragraph*{Theoretical description.}

 
 
\begin{figure}[t]
 \centering
    \includegraphics[width=0.35\linewidth]{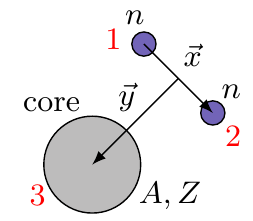}
    \caption{Scaled Jacobi coordinates used in the present work. The three particles are labeled 1 ($n$), 2 ($n$) and 3 (core) for the purpose of defining the pairwise potentials.}
    \label{fig:jac}
\end{figure}

In this work, we describe the three-body $\text{core}+n+n$ system using the hyperspherical formalism~\cite{Zhukov93,Nielsen01}. Within this approach, the wave function for a total angular momentum $j$ is written as
\begin{equation}
  \psi^{j\mu}(\rho,\Omega) = \rho^{-5/2}\sum_{\beta}\chi_{\beta}^{j}(\rho)\mathcal{Y}_{\beta}^{j\mu}(\Omega),
  \label{eq:3bwf}
\end{equation}
where $\rho=\sqrt{x^2+y^2}$ is the hyper-radius defined from the Jacobi-$T$ coordinates in Fig.~\ref{fig:jac}, and $\Omega\equiv\{\alpha,\hat{x},\hat{y}\}$ is introduced for the angular dependence, with $\alpha=\arctan{(x/y)}$ the so-called hyper-angle. The relation between the scaled Jacobi coordinates and the physical distance between the three particles is given by
\begin{equation}
\boldsymbol{x}=\boldsymbol{r}_{x}\sqrt{\frac{1}{2}},~~~~~~~~~~\boldsymbol{y}=\boldsymbol{r}_{y}\sqrt{\frac{2A}{A+2}},
\label{eq:coor}
\end{equation}
where $A$ is the mass number of the core. In Eq.~(\ref{eq:3bwf}), the functions $\mathcal{Y}_{\beta}^{j\mu}(\Omega)$ are expanded in hyperspherical harmonics, which are the analytical eigenfunctions of the hypermomentum operator $\widehat{K}$ \cite{Nielsen01}, and $\beta\equiv\{K,l_x,l_y,l,S_x,J\}$ represents a set of quantum numbers coupled to $j$, with the condition that $n=(K-l_x-l_y)/2$ must be a non-negative integer. 
In this set, $\boldsymbol{l}=\boldsymbol{l}_x+\boldsymbol{l}_y$ is the total orbital angular momentum, $S_x$ is the combined spin of the two neutrons related by the $x$ coordinate, and $\boldsymbol{J}=\boldsymbol{l}+\boldsymbol{S}_x$. With these couplings, and defining $\boldsymbol{I}_c$ as the spin of the core, we have $\boldsymbol{j}=\boldsymbol{J}+\boldsymbol{I}_c$. Note that, in the case that the core is assumed inert, by neglecting its spin we get $J=j$ and the number of components in our wave function expansion~(\ref{eq:3bwf}) is notably reduced. In practice, this expansion is truncated by selecting a $K_{\rm max}$ value, which must be large enough to provide convergence.

The radial functions $\chi_{\beta}^{j}(\rho)$ can be obtained, in general, by solving a set of coupled hyperradial equations defined by the three-body Hamiltonian with pairwise interactions $V_{ij}$ (see, for instance, the details in Ref.~\cite{IJThompson04}). This requires the coupling potentials
\begin{equation}
V_{\beta'\beta}^{j\mu}(\rho)=\left\langle \mathcal{Y}_{\beta }^{j\mu}(\Omega)\Big|V_{12}+V_{13}+V_{23} \Big|\mathcal{Y}_{\beta'}^{ j\mu}(\Omega) \right\rangle_\Omega,
\label{eq:3bcoup}
\end{equation}
where the brakets involve angular and hyperangular integrations. In the case under consideration, $V_{12}=V_{nn}$ and $V_{13}=V_{23}=V_{\text{core-}n}$. In addition to the binary potentials, it is customary to introduce in Eq.~(\ref{eq:3bcoup}) a phenomenological three-body force to account for possible effects that go beyond our strict three-body picture (see, e.g., Refs.~\cite{IJThompson04,MRoGa05,RdDiego10}). This term can be modeled as a diagonal potential $V_{3b}(\rho)\delta_{\beta,\beta'}$, with its hyperradial dependence given by a Gaussian form,
\begin{equation}
    V_{3b}(\rho)=v_{3b}e^{-{\left(\rho/\rho_o\right)}^2}.
    \label{eq:3bforce}
\end{equation}
Typically, the radial parameter $\rho_o$ can be set to 5-6 fm, and the depth $v_{3b}$ is adjusted to fix the energy of the ground state. This choice
is not unique. Another possibility is to introduce some scaling factors in the binary potentials~\cite{Desc03}. 
Here we consider a density-dependent term to modify the central part of the $n$-$n$ interaction, following the prescription in Ref.~\cite{Hagino05},
\begin{equation}
    \tilde{V}_{nn}^{\rm c}(r_x,r_y)=V_{nn}^{\rm c}(r_x)\left(1+\frac{v_o}{1+\exp{[(r_{y}-R_o)/a_o]}}\right).
    \label{eq:nndensdep}
\end{equation}
The additional term is modulated by a Fermi profile for the core nucleus, and here we use $a_o=0.67$ fm and $R_o=1.27A^{1/3}$. At long distances between the core and the center of mass of the two valence neutrons, this term vanishes, and we recover the central part of the bare $n$-$n$ interaction. On the contrary, at short distances it introduces an extra repulsion or attraction depending on the sign of $v_o$, which can be used to shift the energy of the ground state. In the next section we compare the structure properties obtained by using either Eq.~(\ref{eq:3bforce}) or (\ref{eq:nndensdep}) to fix the ground-state energy. 

For simplicity, instead of solving the hyperradial equations with appropriate boundary conditions for bound ($\varepsilon<0$) and continuum ($\varepsilon>0$) states, we expand the radial functions in a discrete basis,
\begin{equation}
    \chi_\beta^j(\rho)=\sum_{i} C_{i\beta}^j U_{i\beta}(\rho),
    \label{eq:PS}
\end{equation}
where the coefficients $C_{i\beta}^j$ can be obtained by diagonalizing the Hamiltonian matrix using $N$ basis functions $\{U_{i\beta}\}$. This approach is typically referred to as as the pseudostate (PS) method~\cite{Tolstikhin97}, which provides a discrete representation of the continuum. Different bases can be employed, but here we adopt the analytical Transformed Harmonic Oscillator (THO) basis~\cite{JCasal13}. As discussed in Refs.~\cite{JCasal13,JCasal14}, these functions have the advantage that their radial extension can be easily tuned to improve the numerical convergence of the ground state, and also to control the concentration of PSs at low excitation energies. 

In particular, we will consider the $B(E1)$ transition strength into the continuum ($\mbox{g.s.} \rightarrow j$), given by
\begin{equation}
    B(E1) = |\langle \text{g.s.} ||\widehat{O}_{E1}||kj\rangle|^2,
    \label{eq:BE1}
\end{equation}
where the dipole operator in Jacobi coordinates is~\cite{JCasal13}
\begin{equation}
    \widehat{O}_{E1} = Ze\sqrt{\frac{2}{A(A+2)}}y Y_{1 M}(\widehat{y}),
    \label{eq:op}
\end{equation}
and $Z$ is the atomic number of the core nucleus. Within our PS representation of the continuum, we obtain discrete $B(E1)$ values for transitions between the bound ground state and dipole states with energy $\varepsilon_k$. In order to construct an energy distribution, we will perform a convolution with Poisson functions preserving the total strength~\cite{MRoGa05,JCasal14}.

\paragraph*{The case of $^{19}$B.} 

Following the procedure described above, we study now the case of $^{19}$B, which has been recently claimed to exhibit a two-neutron halo~\cite{Cook2020}. This nucleus is characterized by a very small (although uncertain) two-neutron separation energy of $S_{2n}=0.089^{+0.560}_{-0.089}$ MeV~\cite{Gaudefroy2012,Wang2017}. In Ref.~\cite{Cook2020}, the value of $S_{2n}=0.5$ MeV was adopted from the best description of the Coulomb dissociation cross section using different calculations. In the present work, we consider various interactions leading to different ground-state properties. For a proper comparison, all our calculations are fixed to produce a ground state at 0.5 MeV below the $^{17}\text{B}+n+n$ threshold. 


For simplicity, our three-body calculations neglect the spin of the $^{17}$B core, so the ground state is characterized by $j^\pi=0^+$. A description beyond this simple picture would require, first, detailed experimental information about the $^{18}$B spectrum, and, second, sophisticated $^{17}\text{B}+n$ interactions leading to a possible splitting of single-particle levels. Furthermore, as shown for the similar case of $^{11}$Li, the precise structure of the ground state is basically sensitive only to the energy of the centroid of the spin-split states \cite{GomezRamos17a,Garrido2020}. Also, the three 1$^-$ states in $^{11}$Li are found to split within a small energy range, producing a quite limited effect on the $B(E1)$ strength \cite{Garrido2002}. For these reasons the role played by the spin of the core in $^{19}$B is not expected to be very relevant. 
Thus, as in Ref.~\cite{Cook2020}, we describe the subsystem with a Woods-Saxon potential
\begin{equation}
V_{n\text{-}^{17}\text{B}}(r) = \left(-V_0+V_{ls} \boldsymbol{l}\cdot\boldsymbol{s}\frac{1}{r}\frac{d}{dr}\right)\frac{1}{1+\exp\left(\frac{r-R}{a}\right)},
\label{vls}
\end{equation}
where $R=1.27A^{1/3}$, $a=0.7$ fm, and the depth parameters of the central and spin-orbit terms are adjusted to reproduce some properties of the unbound $^{18}$B nucleus. The central part is fixed to produce an $s_{1/2}$ virtual state characterized by a scattering length of $a_s=-50$ fm, whereas the spin-orbit potential is determined by the position of a $d_{5/2}$ resonance at 1.1 MeV above the $\text{core}+n$ threshold, which is close to the 1$^-$ state obtained by shell-model calculations~\cite{Spyrou2010}. This yields $V_0=34.3$ MeV and $V_{ls}=34$ MeVfm$^2$.

For the $n$-$n$ subsystem, we employ two different types of interactions. The first one corresponds to realistic, finite-range potentials including central, spin-orbit and tensor terms, which are fixed to reproduce the available $NN$ data. We will show results using two different potentials within this type, the Gogny-Pires-Tourreil (GPT) potential~\cite{GPT}, which has been used with success in several other three-body calculations for $\text{core}+n+n$ nuclei~\cite{Zhukov93,IJThompson04,JCasal13}, and the parametrization Garrido-Fedorov-Jensen (GFJ) specified in Ref.~\cite{Garrido2004b} and employed, for instance,
in Ref.~\cite{Garrido04}. The second type of $n$-$n$ potentials contains just the Gaussian central term,
\begin{equation}
V_{nn}(r_x) = S\exp{[-(r_x/b)^2]},
\label{eq:nngauss}
\end{equation}
whose parameters $(S,b)$ are fixed to yield the known scattering length of $a_s=-15$ fm. In order to explore the sensitivity to finite-range effects, four different sets of parameters are considered : $(S_i,b_i)=(-675.0, 0.4), (-164.0, 0.8), (-24.22, 2.0), (-8.75, 3.2)$. In the small $b$ limit, this Gaussian potential complemented by Eq.~(\ref{eq:nndensdep}) resembles the density-dependent contact pairing interaction used in the three-body calculations of Refs.~\cite{Hagino05,Cook2020}.


We study first the ground-state properties of $^{19}$B and, in particular, its sensitivity to the choice of the $nn$ interaction and the prescription used to fine-tune the two-neutron separation energy. In all cases, convergence has been achieved by including wave-function components in Eq.~(\ref{eq:3bwf})
up to $K_{\rm max}=30$ and $(l_x,l_y)_{max}=3$, 
and $15$ THO basis functions for the Hamiltonian diagonalization. 
The ground-state energy was adjusted to $S_{2n}=0.5$ MeV by fixing either $v_{3b}$ in Eq.~(\ref{eq:3bforce}) or $v_o$ in Eq.~(\ref{eq:nndensdep}). In order to extract the partial-wave content, the ground state was transformed to the Jacobi-$Y$ representation, where the $x$ coordinate connects the core and one of the valence neutrons, using the Raynal-Revai coefficients~\cite{RR70,IJThompson04}.

\begin{table}[t]
    \centering
    \begin{tabular}{cccccc}
      \toprule
                            &  \multicolumn{2}{c}{$V_{3b}$, Eq.~(\ref{eq:3bforce})} &  & \multicolumn{2}{c}{$v_o$, Eq.~(\ref{eq:nndensdep})} \\
      $nn$                  &  $(s_{1/2})^2$ & $(d_{5/2})^2$ &  & $(s_{1/2})^2$ & $(d_{5/2})^2$ \\
      \colrule
      GPT~\cite{GPT}        &  53.2 & 39.2 &  & 54.7 & 38.2 \\
      GFJ~\cite{Garrido2004b} &  56.3 & 35.6 &  & 57.5 & 34.7 \\
      \colrule 
      Gaussian 1                     &  21.5 & 66.1 &  & 24.1 & 68.1 \\   
      Gaussian 2                     &  31.8 & 54.3 &  & 36.8 & 52.1 \\   
      Gaussian 3                     &  52.3 & 41.4 &  & 52.6 & 41.3 \\   
      Gaussian 4                     &  62.7 & 34.2 &  & 57.8 & 37.8 \\   
      \botrule
    \end{tabular}
    \caption{Partial-wave content (\%) for $(s_{1/2})^2$ and $(d_{5/2})^2$ configurations in the ground state of $^{19}$B. The first two lines correspond to the calculations with realistic, finite-range $n$-$n$ interactions. The other lines are the results obtained using the simple central Gaussian potential~(\ref{eq:nngauss}) with parameters $(S_i,b_i)$, $i=1,4$. In all cases, the two-neutron separation energy is fixed to $S_{2n}=0.5$ MeV using the three-body force (Eq.~(\ref{eq:3bforce}), columns 2 and 3) or the density-dependent $n$-$n$ term (Eq.~(\ref{eq:nndensdep}), columns 4 and 5).}
    \label{tab:weights}
\end{table}

\begin{figure}[t]
    \centering
    \includegraphics[width=\linewidth]{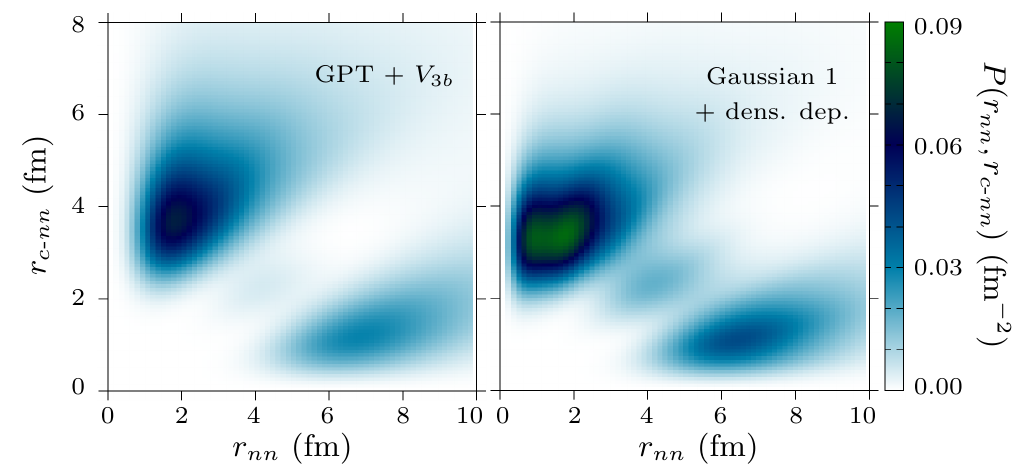}
    \caption{Ground-state probability density of $^{19}$B as a function of $r_{x}\equiv r_{nn}$ and $r_y\equiv r_{c\text{-}nn}$. Left panel: calculations with the GPT $n$-$n$ interaction and $S_{2n}$ fixed to 0.5 MeV with the use of a three-body force. Right panel: calculations with a short-range Gaussian potential (case 1 in Table ~\ref{tab:weights}) and a density-dependent term to fix $S_{2n}$. Both densities are plotted with the same scale.}
    \label{fig:prob}
\end{figure}

In Table~\ref{tab:weights}, we show the weight of the $(s_{1/2})^2$ and $(d_{5/2})^2$ configurations obtained with the different calculations. The results using the two realistic $n$-$n$ interactions yield similar partial-wave contents, with around 55\% occupancy of the $(s_{1/2})$ orbital. It is also worth noting that the particular choice to fix the ground-state energy in this case, via Eq.~(\ref{eq:3bforce}) or Eq.~(\ref{eq:nndensdep}), has little effect on the final weights. When a more sophisticated $n$-$n$ interaction, such as AV18~\cite{av18}, is used, the results are very similar to the ones shown in the upper part of  Table~\ref{tab:weights}. This stability is related to a correct description of the low-energy nucleon-nucleon properties more than to the presence or not of a short-distance infinite repulsion. The
choice of smooth potentials, such as the ones used in this work permits, however, to speed up the calculations. 
On the contrary, the calculations using the simple Gaussian potentials provide very different results depending on the range of the interaction. In particular, as the $n$-$n$ potential becomes narrower, the $s$-wave content decreases. For a range smaller than $\approx 1$ fm, we obtain an inversion of the weights, such that the $(d_{5/2})^2$ configuration dominates. This resembles the results presented in Ref.~\cite{Cook2020}, where a value of 35\% for the $s_{1/2}$ orbital was reported. Note that with this kind of simple interactions, the probabilities are more sensitive to the particular choice used to fix the ground-state energy.

The dramatic difference in $s$-wave content between calculations using a realistic $n$-$n$ interaction and those with a contact-like potential leads to ground states characterized by distinct radial probabilities. In Fig.~\ref{fig:prob} we present the wave-function probability as a function of $r_{x}\equiv r_{nn}$ and $r_y\equiv r_{c\text{-}nn}$ for two of the present calculations: (i) left panel, the GPT interaction plus the three-body force fixing the ground-state energy (GPT row, columns 2 and 3, in Table~\ref{tab:weights}), and (ii) right panel, the contact-like Gaussian 1 potential plus the density-dependent term (Gaussian 1 row, columns 4 and 5, in Table~\ref{tab:weights}). Both scenarios give rise to a maximum corresponding to the two neutrons close to each other at some distance from the core, which is typically referred to as the dineutron peak. However, the latter produces a more localized wave function, which is a consequence of the smaller $s$-wave content. It is also apparent that the valence neutrons explore shorter relative distances than the ones available when considering the realistic $n$-$n$ interaction. These features are reflected by the rms distances $\langle r_x^2\rangle^{1/2}$ and $\langle r_y^2\rangle^{1/2}$, which take the values of 7.28 and 5.01 fm, respectively, in the realistic case, and 5.96 fm and 3.90 fm in the contact-like case.


The ground-state properties and, in particular, the $s_{1/2}$ content of the wave function have an impact on the low-lying $E1$ response of halo nuclei. The sum rule for dipole transitions is totally determined by the ground state~\cite{JCasal13}. As discussed, the choice of the $n$-$n$ interaction may produce important differences in terms of partial-wave content. Here we explore its effect on the $B(E1)$ distribution for $^{19}$B. This requires the construction of dipole (1$^-$) states above the three-body threshold, that we generate using 35 THO basis functions in order to increase the pseudostate level density. Then, we compute discrete $B(E1)$ values, Eqs.~(\ref{eq:BE1}) and (\ref{eq:op}), which are smeared using Poisson functions to get the corresponding $dB(E1)/d\varepsilon$ 
distribution.

\begin{figure}[t]
    \centering
    \includegraphics[width=\linewidth]{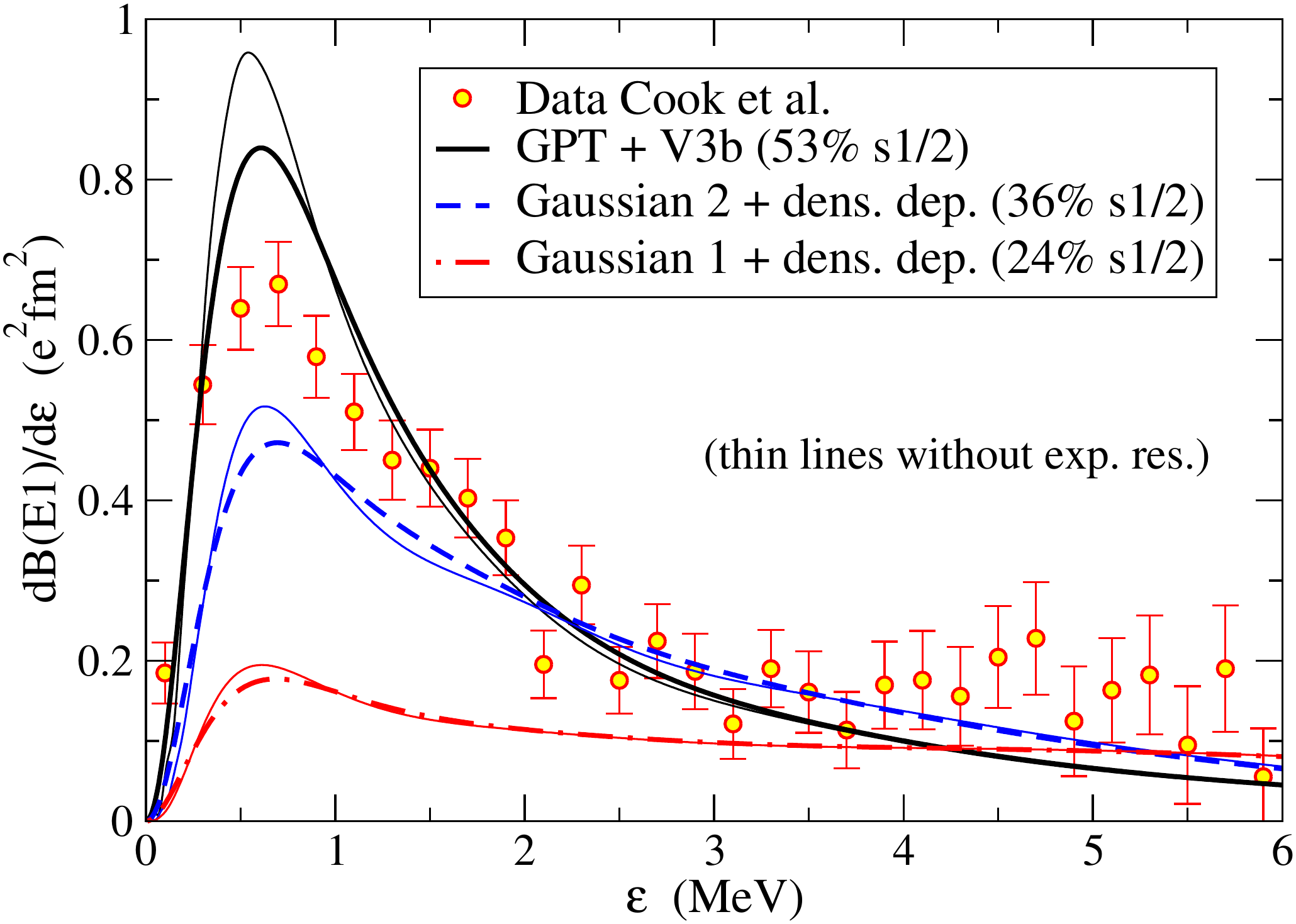}
    \caption{$B(E1)$ distribution for $^{19}$B. Calculations correspond to the results with the GPT interaction and the three-body force (solid black line), and the simple Gaussian 1 (dot-dashed red line) and 2 (dashed blue line) complemented by the density-dependent term. The curves have been convoluted with the experimental resolution reported Ref.~\cite{Cook2020} as a Gaussian distribution of width $\sigma(\varepsilon)=0.25\varepsilon^{0.53}$ MeV. The thin lines are the results before the convolution.}
    \label{fig:BE1}
\end{figure}

Our results are shown in Fig.~\ref{fig:BE1} together with the experimental data from Ref.~\cite{Cook2020}. Again, we focus on the case of the GPT interaction plus the three-body force (thick black line) and the contact-like Gaussian 1 potential plus the density-dependent term (dot-dashed red line). These curves include the convolution with the experimental resolution reported in Ref.~\cite{Cook2020}, but for completeness we include also the purely theoretical results (thin lines). From this figure it is clear that the $B(E1)$ distribution obtained using the realistic, finite-range GPT interaction, which provides 53.2\% of $s$-waves in the ground state, gives a better description of the data. Although the calculations tend to slightly overestimate the data around the low-energy peak, the calculated total strength up to 6 MeV is $1.53$ e$^2$fm$^2$, compatible with the experimental value of $B(E1)=1.64\pm0.06({\rm stat})\pm0.12({\rm sys})$ e$^2$fm$^2$. In contrast, with our contact-like interaction, which gave only 24.1\% of $s$-waves in the ground state, the data are strongly underestimated. We show also the results considering instead the Gaussian 2 potential (dashed blue line), which was found to produce an $s$-wave content of 36.8\%. In that case, the $B(E1)$ strength up to 6 MeV is 1.26 e$^2$fm$^2$, i.e., about 20\% smaller than the value obtained with our tensor interaction, and the theoretical curve slightly underestimates the data, similarly to the results in Ref.~\cite{Cook2020}. 

It is important to remember that the $^{17}$B-$n$ interaction has been constructed just following Ref.~\cite{Cook2020}. In particular, the $s$-wave potential has been designed to reproduce the scattering length $a_s=-50$ fm, given in Ref.~\cite{Cook2020} as the one reproducing better the experimental cross section. However, there is no other reason to consider this value of $a_s$ as the correct one (see, e.g., the calculations in Ref.~\cite{Hiyama19} using different $a_s$ values). Also, we can see in Ref.~\cite{Cook2020} 
that a smaller value of $|a_s|$ clearly lowers the peak in the $B(E1)$ distribution. Therefore, the overestimation of the peak height shown by
our calculation in Fig.~\ref{fig:BE1} can be an indication that, due to the use of finite-range potentials, some tuning of the $^{17}$B-$n$ interaction could be needed.  Given the uncertainties in both the $S_{2n}$ and in the low-lying unbound spectrum of $^{18}$B, we prefer to keep the values adopted in Ref.~\cite{Cook2020} for an easy comparison with the previous calculations.

Finally, note that the present calculations with a simple central Gaussian dependence of the $n$-$n$ interaction are not fully equivalent to the contact potential in Refs.~\cite{Hagino05,Cook2020}. First, we simulate a $\delta(r_{nn})$ dependence, whereas in Refs.~\cite{Hagino05,Cook2020} a dependence on the vector $\bm{r}_{nn}$ is considered, i.e., $\delta(\bm{r}_{nn})$. And second, working in coordinate space we do not introduce any energy cutoff in the potential spectrum. Our simple Gaussian potentials are constructed to reproduce just the $n$-$n$ scattering length, whereas in Refs.~\cite{Hagino05,Cook2020}, as shown in Fig.~1 of Ref.~\cite{Esbensen1997}, the energy cutoff is used to reproduce also reasonably well the low-energy phase shifts. This may explain why our three-body calculations with the narrow Gaussian potentials yield even smaller $s$-wave contents than in Ref.~\cite{Cook2020} as the range of the potential decreases. However, from the comparison between our results and the experimental $B(E1)$, we stress that a model with a finite-range $n$-$n$ interaction leading to $\approx 55\%$ of $s$-waves in the ground state of $^{19}$B is fully compatible with the available data.

  


\paragraph*{Summary and  conclusions.}

In this work the three-body model in Ref.~\cite{JCasal13} has been used to describe the two-neutron halo nucleus $^{19}$B ($^{17}$B+$n$+$n$). 
First, the structure of $^{19}$B and the $B(E1)$ transition strength have been investigated by use of finite-range two-body interactions. Following Ref.~\cite{Cook2020}, the spin of the core is neglected, and the core-neutron potential is taken as a Woods-Saxon shape, whose parameters are adjusted to the scattering length $a_s=-50$ fm and the $d_{5/2}$ resonance energy predicted by shell-model calculations \cite{Spyrou2010}. For the $n$-$n$ interaction two different tensor potentials have been considered: the GPT potential \cite{GPT} and the GFJ parametrization in Ref.~\cite{Garrido2004b}. 
Two different procedures have been used to adjust the two-neutron separation energy: (i) a Gaussian three-body force 
and (ii) a density-dependent term in the central part of the $n$-$n$ potential, as in Ref.~\cite{Esbensen1997}.

The results obtained are stable, with not relevant variations depending on the realistic $n$-$n$ interaction and on the method used to fine tune the three-body energy. We have obtained a dominant contribution of the $(s_{1/2})^2$ component, which provides about $53$-$57$\% of the norm, whereas the weight of the $(d_{5/2})^2$ component moves within the $35$-$39$\% range. The computed $dB(E1)/d\varepsilon$ agrees rather well with the available data after convolution with the experimental resolution. 
The integrated $B(E1)$ strength (up to 6 MeV) is $1.53$ e$^2$fm$^2$, also in good agreement with the experimental value~\cite{Cook2020}.

The second goal has been to investigate the effects due to the range of the $n$-$n$ interaction. 
In Ref.~\cite{Cook2020}, where a density-dependent contact $n$-$n$ interaction was used, the predicted weights for the $(s_{1/2})^2$ and $(d_{5/2})^2$ components were, respectively, of about 35\% and 56\%, essentially the opposite to our present calculations. Having this in mind, we have considered four different cases where the $n$-$n$ interaction is modeled as a simple central Gaussian potential adjusted to reproduce the correct $n$-$n$ scattering length. We have found that a smaller Gaussian range implies also a reduction (increase) in the $s$-wave ($d$-wave) content, in such a way that values consistent with Ref.~\cite{Cook2020} are obtained for Gaussian ranges below 1 fm. Furthermore, for ranges of about a few femtometers, the weights of the $(s_{1/2})^2$ and $(d_{5/2})^2$ components are consistent with the results obtained with the realistic $n$-$n$ potentials. This result explains as well the too low $s$-wave content found in Ref.~\cite{Esbensen1997} for $^{11}$Li. As a consequence of the decrease in the $s$-wave content, the system becomes more compact, reducing by about 20\% the $n$-$n$ and core-($nn$) distances. 
This leads to a systematic underestimation of the $B(E1)$ distribution. 

To summarize, we conclude that the three-body model appears as an appropriate description of $^{19}$B, consistent with the two-neutron halo structure reported in Ref.~\cite{Cook2020}. A good agreement with the available experimental data is found. Although the results are, to a large extent, independent of the short-distance details of the two-body potentials, it is, however, important to maintain the finite-range character of the two-body interactions. 

\section*{Acknowledgements}

The authors are grateful to K.~Hagino, K.~J.~Cook, and M. Rodríguez-Gallardo for useful discussions. This work has been partially supported by SID funds 2019 (Università degli Studi di Padova, Italy) under Project No.~CASA\_SID19\_01, by the Spanish Ministry of Science, Innovation and University MCIU/AEI/FEDER,UE (Spain) under Contract No. PGC2018-093636-B-I00, and by the European Union's Horizon 2020 research and innovation program under grant agreement No.~654002 (ENSAR2).

\bibliography{bibfile}

\begin{thebibliography}{32}%
\makeatletter
\providecommand \@ifxundefined [1]{%
 \@ifx{#1\undefined}
}%
\providecommand \@ifnum [1]{%
 \ifnum #1\expandafter \@firstoftwo
 \else \expandafter \@secondoftwo
 \fi
}%
\providecommand \@ifx [1]{%
 \ifx #1\expandafter \@firstoftwo
 \else \expandafter \@secondoftwo
 \fi
}%
\providecommand \natexlab [1]{#1}%
\providecommand \enquote  [1]{``#1''}%
\providecommand \bibnamefont  [1]{#1}%
\providecommand \bibfnamefont [1]{#1}%
\providecommand \citenamefont [1]{#1}%
\providecommand \href@noop [0]{\@secondoftwo}%
\providecommand \href [0]{\begingroup \@sanitize@url \@href}%
\providecommand \@href[1]{\@@startlink{#1}\@@href}%
\providecommand \@@href[1]{\endgroup#1\@@endlink}%
\providecommand \@sanitize@url [0]{\catcode `\\12\catcode `\$12\catcode
  `\&12\catcode `\#12\catcode `\^12\catcode `\_12\catcode `\%12\relax}%
\providecommand \@@startlink[1]{}%
\providecommand \@@endlink[0]{}%
\providecommand \url  [0]{\begingroup\@sanitize@url \@url }%
\providecommand \@url [1]{\endgroup\@href {#1}{\urlprefix }}%
\providecommand \urlprefix  [0]{URL }%
\providecommand \Eprint [0]{\href }%
\providecommand \doibase [0]{http://dx.doi.org/}%
\providecommand \selectlanguage [0]{\@gobble}%
\providecommand \bibinfo  [0]{\@secondoftwo}%
\providecommand \bibfield  [0]{\@secondoftwo}%
\providecommand \translation [1]{[#1]}%
\providecommand \BibitemOpen [0]{}%
\providecommand \bibitemStop [0]{}%
\providecommand \bibitemNoStop [0]{.\EOS\space}%
\providecommand \EOS [0]{\spacefactor3000\relax}%
\providecommand \BibitemShut  [1]{\csname bibitem#1\endcsname}%
\let\auto@bib@innerbib\@empty
\bibitem [{\citenamefont {Tanihata}\ \emph {et~al.}(1985)\citenamefont
  {Tanihata} \emph {et~al.}}]{Tanihata1985}%
  \BibitemOpen
  \bibfield  {author} {\bibinfo {author} {\bibfnamefont {I.}~\bibnamefont
  {Tanihata}} \emph {et~al.},\ }\href {\doibase 10.1103/PhysRevLett.55.2676}
  {\bibfield  {journal} {\bibinfo  {journal} {Phys. Rev. Lett.}\ }\textbf
  {\bibinfo {volume} {55}},\ \bibinfo {pages} {2676} (\bibinfo {year}
  {1985})}\BibitemShut {NoStop}%
\bibitem [{\citenamefont {Riisager}(1994)}]{Riisager1994}%
  \BibitemOpen
  \bibfield  {author} {\bibinfo {author} {\bibfnamefont {K.}~\bibnamefont
  {Riisager}},\ }\href {\doibase 10.1103/RevModPhys.66.1105} {\bibfield
  {journal} {\bibinfo  {journal} {Rev. Mod. Phys.}\ }\textbf {\bibinfo {volume}
  {66}},\ \bibinfo {pages} {1105} (\bibinfo {year} {1994})}\BibitemShut
  {NoStop}%
\bibitem [{\citenamefont {Hansen}\ and\ \citenamefont
  {Jonson}(1987)}]{Hansen1987}%
  \BibitemOpen
  \bibfield  {author} {\bibinfo {author} {\bibfnamefont {P.}~\bibnamefont
  {Hansen}}\ and\ \bibinfo {author} {\bibfnamefont {J.}~\bibnamefont
  {Jonson}},\ }\href {\doibase 10.1209/0295-5075/4/4/005} {\bibfield  {journal}
  {\bibinfo  {journal} {Europhys. Lett.}\ }\textbf {\bibinfo {volume} {4}},\
  \bibinfo {pages} {409} (\bibinfo {year} {1987})}\BibitemShut {NoStop}%
\bibitem [{\citenamefont {Zhukov}\ \emph {et~al.}(1993)\citenamefont {Zhukov},
  \citenamefont {Danilin}, \citenamefont {Fedorov}, \citenamefont {Bang},
  \citenamefont {Thompson},\ and\ \citenamefont {Vaagen}}]{Zhukov93}%
  \BibitemOpen
  \bibfield  {author} {\bibinfo {author} {\bibfnamefont {M.}~\bibnamefont
  {Zhukov}}, \bibinfo {author} {\bibfnamefont {B.}~\bibnamefont {Danilin}},
  \bibinfo {author} {\bibfnamefont {D.}~\bibnamefont {Fedorov}}, \bibinfo
  {author} {\bibfnamefont {J.}~\bibnamefont {Bang}}, \bibinfo {author}
  {\bibfnamefont {I.}~\bibnamefont {Thompson}}, \ and\ \bibinfo {author}
  {\bibfnamefont {J.}~\bibnamefont {Vaagen}},\ }\href {\doibase
  10.1016/0370-1573(93)90141-Y} {\bibfield  {journal} {\bibinfo  {journal}
  {Phys. Rep.}\ }\textbf {\bibinfo {volume} {231}},\ \bibinfo {pages} {151}
  (\bibinfo {year} {1993})}\BibitemShut {NoStop}%
\bibitem [{\citenamefont {Thompson}\ \emph {et~al.}(2004)\citenamefont
  {Thompson}, \citenamefont {Nunes},\ and\ \citenamefont
  {Danilin}}]{IJThompson04}%
  \BibitemOpen
  \bibfield  {author} {\bibinfo {author} {\bibfnamefont {I.}~\bibnamefont
  {Thompson}}, \bibinfo {author} {\bibfnamefont {F.}~\bibnamefont {Nunes}}, \
  and\ \bibinfo {author} {\bibfnamefont {B.}~\bibnamefont {Danilin}},\ }\href
  {\doibase 10.1016/j.cpc.2004.03.007} {\bibfield  {journal} {\bibinfo
  {journal} {Comp. Phys. Commun.}\ }\textbf {\bibinfo {volume} {161}},\
  \bibinfo {pages} {87} (\bibinfo {year} {2004})}\BibitemShut {NoStop}%
\bibitem [{\citenamefont {Garrido}\ \emph
  {et~al.}(2004{\natexlab{a}})\citenamefont {Garrido}, \citenamefont
  {Fedorov},\ and\ \citenamefont {Jensen}}]{Garrido04}%
  \BibitemOpen
  \bibfield  {author} {\bibinfo {author} {\bibfnamefont {E.}~\bibnamefont
  {Garrido}}, \bibinfo {author} {\bibfnamefont {D.~V.}\ \bibnamefont
  {Fedorov}}, \ and\ \bibinfo {author} {\bibfnamefont {A.~S.}\ \bibnamefont
  {Jensen}},\ }\href {\doibase 10.1016/j.nuclphysa.2003.12.016} {\bibfield
  {journal} {\bibinfo  {journal} {Nucl. Phys. A}\ }\textbf {\bibinfo {volume}
  {733}},\ \bibinfo {pages} {85} (\bibinfo {year}
  {2004}{\natexlab{a}})}\BibitemShut {NoStop}%
\bibitem [{\citenamefont {Wiringa}\ \emph {et~al.}(1995)\citenamefont
  {Wiringa}, \citenamefont {Stoks},\ and\ \citenamefont {Schiavilla}}]{av18}%
  \BibitemOpen
  \bibfield  {author} {\bibinfo {author} {\bibfnamefont {R.~B.}\ \bibnamefont
  {Wiringa}}, \bibinfo {author} {\bibfnamefont {V.~G.~J.}\ \bibnamefont
  {Stoks}}, \ and\ \bibinfo {author} {\bibfnamefont {R.}~\bibnamefont
  {Schiavilla}},\ }\href {\doibase 10.1103/PhysRevC.51.38} {\bibfield
  {journal} {\bibinfo  {journal} {Phys. Rev. C}\ }\textbf {\bibinfo {volume}
  {51}},\ \bibinfo {pages} {38} (\bibinfo {year} {1995})}\BibitemShut {NoStop}%
\bibitem [{\citenamefont {Esbensen}\ \emph {et~al.}(1997)\citenamefont
  {Esbensen}, \citenamefont {Bertsch},\ and\ \citenamefont
  {Hencken}}]{Esbensen1997}%
  \BibitemOpen
  \bibfield  {author} {\bibinfo {author} {\bibfnamefont {H.}~\bibnamefont
  {Esbensen}}, \bibinfo {author} {\bibfnamefont {G.~F.}\ \bibnamefont
  {Bertsch}}, \ and\ \bibinfo {author} {\bibfnamefont {K.}~\bibnamefont
  {Hencken}},\ }\href {\doibase 10.1103/PhysRevC.56.3054} {\bibfield  {journal}
  {\bibinfo  {journal} {Phys. Rev. C}\ }\textbf {\bibinfo {volume} {56}},\
  \bibinfo {pages} {3054} (\bibinfo {year} {1997})}\BibitemShut {NoStop}%
\bibitem [{\citenamefont {Hagino}\ and\ \citenamefont
  {Sagawa}(2005)}]{Hagino05}%
  \BibitemOpen
  \bibfield  {author} {\bibinfo {author} {\bibfnamefont {K.}~\bibnamefont
  {Hagino}}\ and\ \bibinfo {author} {\bibfnamefont {H.}~\bibnamefont
  {Sagawa}},\ }\href {\doibase 10.1103/PhysRevC.72.044321} {\bibfield
  {journal} {\bibinfo  {journal} {Phys. Rev. C}\ }\textbf {\bibinfo {volume}
  {72}},\ \bibinfo {pages} {044321} (\bibinfo {year} {2005})}\BibitemShut
  {NoStop}%
\bibitem [{\citenamefont {Young}\ \emph {et~al.}(1994)\citenamefont {Young}
  \emph {et~al.}}]{Young1994}%
  \BibitemOpen
  \bibfield  {author} {\bibinfo {author} {\bibfnamefont {B.~M.}\ \bibnamefont
  {Young}} \emph {et~al.},\ }\href {\doibase 10.1103/PhysRevC.49.279}
  {\bibfield  {journal} {\bibinfo  {journal} {Phys. Rev. C}\ }\textbf {\bibinfo
  {volume} {49}},\ \bibinfo {pages} {279} (\bibinfo {year} {1994})}\BibitemShut
  {NoStop}%
\bibitem [{\citenamefont {Zinser}\ \emph {et~al.}(1997)\citenamefont {Zinser}
  \emph {et~al.}}]{Zinser1997}%
  \BibitemOpen
  \bibfield  {author} {\bibinfo {author} {\bibfnamefont {M.}~\bibnamefont
  {Zinser}} \emph {et~al.},\ }\href {\doibase 10.1016/S0375-9474(97)00134-6}
  {\bibfield  {journal} {\bibinfo  {journal} {Nucl. Phys. A}\ }\textbf
  {\bibinfo {volume} {619}},\ \bibinfo {pages} {151} (\bibinfo {year}
  {1997})}\BibitemShut {NoStop}%
\bibitem [{\citenamefont {Garrido}\ \emph {et~al.}(1997)\citenamefont
  {Garrido}, \citenamefont {Fedorov},\ and\ \citenamefont
  {Jensen}}]{Garrido1997}%
  \BibitemOpen
  \bibfield  {author} {\bibinfo {author} {\bibfnamefont {E.}~\bibnamefont
  {Garrido}}, \bibinfo {author} {\bibfnamefont {D.~V.}\ \bibnamefont
  {Fedorov}}, \ and\ \bibinfo {author} {\bibfnamefont {A.~S.}\ \bibnamefont
  {Jensen}},\ }\href {\doibase 10.1103/PhysRevC.55.1327} {\bibfield  {journal}
  {\bibinfo  {journal} {Phys. Rev. C}\ }\textbf {\bibinfo {volume} {55}},\
  \bibinfo {pages} {1327} (\bibinfo {year} {1997})}\BibitemShut {NoStop}%
\bibitem [{\citenamefont {Casal}\ \emph {et~al.}(2017)\citenamefont {Casal},
  \citenamefont {G{\'{o}}mez-Ramos},\ and\ \citenamefont {Moro}}]{Casal2017a}%
  \BibitemOpen
  \bibfield  {author} {\bibinfo {author} {\bibfnamefont {J.}~\bibnamefont
  {Casal}}, \bibinfo {author} {\bibfnamefont {M.}~\bibnamefont
  {G{\'{o}}mez-Ramos}}, \ and\ \bibinfo {author} {\bibfnamefont {A.~M.}\
  \bibnamefont {Moro}},\ }\href {\doibase 10.1016/j.physletb.2017.02.017}
  {\bibfield  {journal} {\bibinfo  {journal} {Phys. Lett. B}\ }\textbf
  {\bibinfo {volume} {767}},\ \bibinfo {pages} {307} (\bibinfo {year}
  {2017})}\BibitemShut {NoStop}%
\bibitem [{\citenamefont {Gómez-Ramos}\ \emph {et~al.}(2017)\citenamefont
  {Gómez-Ramos}, \citenamefont {Casal},\ and\ \citenamefont
  {Moro}}]{GomezRamos17a}%
  \BibitemOpen
  \bibfield  {author} {\bibinfo {author} {\bibfnamefont {M.}~\bibnamefont
  {Gómez-Ramos}}, \bibinfo {author} {\bibfnamefont {J.}~\bibnamefont {Casal}},
  \ and\ \bibinfo {author} {\bibfnamefont {A.}~\bibnamefont {Moro}},\ }\href
  {\doibase 10.1016/j.physletb.2017.06.023} {\bibfield  {journal} {\bibinfo
  {journal} {Phys. Lett. B}\ }\textbf {\bibinfo {volume} {772}},\ \bibinfo
  {pages} {115} (\bibinfo {year} {2017})}\BibitemShut {NoStop}%
\bibitem [{\citenamefont {Oishi}\ \emph {et~al.}(2017)\citenamefont {Oishi},
  \citenamefont {Kortelainen},\ and\ \citenamefont {Pastore}}]{Oishi2017}%
  \BibitemOpen
  \bibfield  {author} {\bibinfo {author} {\bibfnamefont {T.}~\bibnamefont
  {Oishi}}, \bibinfo {author} {\bibfnamefont {M.}~\bibnamefont {Kortelainen}},
  \ and\ \bibinfo {author} {\bibfnamefont {A.}~\bibnamefont {Pastore}},\ }\href
  {\doibase 10.1103/PhysRevC.96.044327} {\bibfield  {journal} {\bibinfo
  {journal} {Phys. Rev. C}\ }\textbf {\bibinfo {volume} {96}},\ \bibinfo
  {pages} {044327} (\bibinfo {year} {2017})}\BibitemShut {NoStop}%
\bibitem [{\citenamefont {Cook}\ \emph {et~al.}(2020)\citenamefont {Cook} \emph
  {et~al.}}]{Cook2020}%
  \BibitemOpen
  \bibfield  {author} {\bibinfo {author} {\bibfnamefont {K.~J.}\ \bibnamefont
  {Cook}} \emph {et~al.},\ }\href {\doibase 10.1103/PhysRevLett.124.212503}
  {\bibfield  {journal} {\bibinfo  {journal} {Phys. Rev. Lett.}\ }\textbf
  {\bibinfo {volume} {124}},\ \bibinfo {pages} {212503} (\bibinfo {year}
  {2020})}\BibitemShut {NoStop}%
\bibitem [{\citenamefont {Nielsen}\ \emph {et~al.}(2001)\citenamefont
  {Nielsen}, \citenamefont {Fedorov}, \citenamefont {Jensen},\ and\
  \citenamefont {Garrido}}]{Nielsen01}%
  \BibitemOpen
  \bibfield  {author} {\bibinfo {author} {\bibfnamefont {E.}~\bibnamefont
  {Nielsen}}, \bibinfo {author} {\bibfnamefont {D.~V.}\ \bibnamefont
  {Fedorov}}, \bibinfo {author} {\bibfnamefont {A.~S.}\ \bibnamefont {Jensen}},
  \ and\ \bibinfo {author} {\bibfnamefont {E.}~\bibnamefont {Garrido}},\ }\href
  {\doibase 10.1016/S0370-1573(00)00107-1} {\bibfield  {journal} {\bibinfo
  {journal} {Phys. Rep.}\ }\textbf {\bibinfo {volume} {347}},\ \bibinfo {pages}
  {373} (\bibinfo {year} {2001})}\BibitemShut {NoStop}%
\bibitem [{\citenamefont {Rodr\'{\i}guez-Gallardo}\ \emph
  {et~al.}(2005)\citenamefont {Rodr\'{\i}guez-Gallardo}, \citenamefont {Arias},
  \citenamefont {G\'omez-Camacho}, \citenamefont {Moro}, \citenamefont
  {Thompson},\ and\ \citenamefont {Tostevin}}]{MRoGa05}%
  \BibitemOpen
  \bibfield  {author} {\bibinfo {author} {\bibfnamefont {M.}~\bibnamefont
  {Rodr\'{\i}guez-Gallardo}}, \bibinfo {author} {\bibfnamefont {J.~M.}\
  \bibnamefont {Arias}}, \bibinfo {author} {\bibfnamefont {J.}~\bibnamefont
  {G\'omez-Camacho}}, \bibinfo {author} {\bibfnamefont {A.~M.}\ \bibnamefont
  {Moro}}, \bibinfo {author} {\bibfnamefont {I.~J.}\ \bibnamefont {Thompson}},
  \ and\ \bibinfo {author} {\bibfnamefont {J.~A.}\ \bibnamefont {Tostevin}},\
  }\href {\doibase 10.1103/PhysRevC.72.024007} {\bibfield  {journal} {\bibinfo
  {journal} {Phys. Rev. C}\ }\textbf {\bibinfo {volume} {72}},\ \bibinfo
  {pages} {024007} (\bibinfo {year} {2005})}\BibitemShut {NoStop}%
\bibitem [{\citenamefont {de~Diego}\ \emph {et~al.}(2010)\citenamefont
  {de~Diego}, \citenamefont {Garrido}, \citenamefont {Fedorov},\ and\
  \citenamefont {Jensen}}]{RdDiego10}%
  \BibitemOpen
  \bibfield  {author} {\bibinfo {author} {\bibfnamefont {R.}~\bibnamefont
  {de~Diego}}, \bibinfo {author} {\bibfnamefont {E.}~\bibnamefont {Garrido}},
  \bibinfo {author} {\bibfnamefont {D.~V.}\ \bibnamefont {Fedorov}}, \ and\
  \bibinfo {author} {\bibfnamefont {A.~S.}\ \bibnamefont {Jensen}},\
  }\href@noop {} {\bibfield  {journal} {\bibinfo  {journal} {Europhys. Lett.}\
  }\textbf {\bibinfo {volume} {90}},\ \bibinfo {pages} {52001} (\bibinfo {year}
  {2010})}\BibitemShut {NoStop}%
\bibitem [{\citenamefont {Descouvemont}\ \emph {et~al.}(2003)\citenamefont
  {Descouvemont}, \citenamefont {Daniel},\ and\ \citenamefont {Baye}}]{Desc03}%
  \BibitemOpen
  \bibfield  {author} {\bibinfo {author} {\bibfnamefont {P.}~\bibnamefont
  {Descouvemont}}, \bibinfo {author} {\bibfnamefont {C.}~\bibnamefont
  {Daniel}}, \ and\ \bibinfo {author} {\bibfnamefont {D.}~\bibnamefont
  {Baye}},\ }\href {\doibase 10.1103/PhysRevC.67.044309} {\bibfield  {journal}
  {\bibinfo  {journal} {Phys. Rev. C}\ }\textbf {\bibinfo {volume} {67}},\
  \bibinfo {pages} {044309} (\bibinfo {year} {2003})}\BibitemShut {NoStop}%
\bibitem [{\citenamefont {Tolstikhin}\ \emph {et~al.}(1997)\citenamefont
  {Tolstikhin}, \citenamefont {Ostrovsky},\ and\ \citenamefont
  {Nakamura}}]{Tolstikhin97}%
  \BibitemOpen
  \bibfield  {author} {\bibinfo {author} {\bibfnamefont {O.~I.}\ \bibnamefont
  {Tolstikhin}}, \bibinfo {author} {\bibfnamefont {V.~N.}\ \bibnamefont
  {Ostrovsky}}, \ and\ \bibinfo {author} {\bibfnamefont {H.}~\bibnamefont
  {Nakamura}},\ }\href {\doibase 10.1103/PhysRevLett.79.2026} {\bibfield
  {journal} {\bibinfo  {journal} {Phys. Rev. Lett.}\ }\textbf {\bibinfo
  {volume} {79}},\ \bibinfo {pages} {2026} (\bibinfo {year}
  {1997})}\BibitemShut {NoStop}%
\bibitem [{\citenamefont {Casal}\ \emph {et~al.}(2013)\citenamefont {Casal},
  \citenamefont {Rodr\'{\i}guez-Gallardo},\ and\ \citenamefont
  {Arias}}]{JCasal13}%
  \BibitemOpen
  \bibfield  {author} {\bibinfo {author} {\bibfnamefont {J.}~\bibnamefont
  {Casal}}, \bibinfo {author} {\bibfnamefont {M.}~\bibnamefont
  {Rodr\'{\i}guez-Gallardo}}, \ and\ \bibinfo {author} {\bibfnamefont {J.~M.}\
  \bibnamefont {Arias}},\ }\href {\doibase 10.1103/PhysRevC.88.014327}
  {\bibfield  {journal} {\bibinfo  {journal} {Phys. Rev. C}\ }\textbf {\bibinfo
  {volume} {88}},\ \bibinfo {pages} {014327} (\bibinfo {year}
  {2013})}\BibitemShut {NoStop}%
\bibitem [{\citenamefont {Casal}\ \emph {et~al.}(2014)\citenamefont {Casal},
  \citenamefont {Rodr\'{\i}guez-Gallardo}, \citenamefont {Arias},\ and\
  \citenamefont {Thompson}}]{JCasal14}%
  \BibitemOpen
  \bibfield  {author} {\bibinfo {author} {\bibfnamefont {J.}~\bibnamefont
  {Casal}}, \bibinfo {author} {\bibfnamefont {M.}~\bibnamefont
  {Rodr\'{\i}guez-Gallardo}}, \bibinfo {author} {\bibfnamefont {J.~M.}\
  \bibnamefont {Arias}}, \ and\ \bibinfo {author} {\bibfnamefont {I.~J.}\
  \bibnamefont {Thompson}},\ }\href {\doibase 10.1103/PhysRevC.90.044304}
  {\bibfield  {journal} {\bibinfo  {journal} {Phys. Rev. C}\ }\textbf {\bibinfo
  {volume} {90}},\ \bibinfo {pages} {044304} (\bibinfo {year}
  {2014})}\BibitemShut {NoStop}%
\bibitem [{\citenamefont {Gaudefroy}\ \emph {et~al.}(2012)\citenamefont
  {Gaudefroy} \emph {et~al.}}]{Gaudefroy2012}%
  \BibitemOpen
  \bibfield  {author} {\bibinfo {author} {\bibfnamefont {L.}~\bibnamefont
  {Gaudefroy}} \emph {et~al.},\ }\href {\doibase
  10.1103/PhysRevLett.109.202503} {\bibfield  {journal} {\bibinfo  {journal}
  {Phys. Rev. Lett.}\ }\textbf {\bibinfo {volume} {109}},\ \bibinfo {pages}
  {202503} (\bibinfo {year} {2012})}\BibitemShut {NoStop}%
\bibitem [{\citenamefont {Wang}\ \emph {et~al.}(2017)\citenamefont {Wang},
  \citenamefont {Audi}, \citenamefont {Kondev}, \citenamefont {Huang},
  \citenamefont {Naimi},\ and\ \citenamefont {Xu}}]{Wang2017}%
  \BibitemOpen
  \bibfield  {author} {\bibinfo {author} {\bibfnamefont {M.}~\bibnamefont
  {Wang}}, \bibinfo {author} {\bibfnamefont {G.}~\bibnamefont {Audi}}, \bibinfo
  {author} {\bibfnamefont {F.~G.}\ \bibnamefont {Kondev}}, \bibinfo {author}
  {\bibfnamefont {W.}~\bibnamefont {Huang}}, \bibinfo {author} {\bibfnamefont
  {S.}~\bibnamefont {Naimi}}, \ and\ \bibinfo {author} {\bibfnamefont
  {X.}~\bibnamefont {Xu}},\ }\href {\doibase 10.1088/1674-1137/41/3/030003}
  {\bibfield  {journal} {\bibinfo  {journal} {Chin. Phys. C}\ }\textbf
  {\bibinfo {volume} {41}},\ \bibinfo {pages} {030003} (\bibinfo {year}
  {2017})}\BibitemShut {NoStop}%
\bibitem [{\citenamefont {Garrido}\ and\ \citenamefont
  {Jensen}(2020)}]{Garrido2020}%
  \BibitemOpen
  \bibfield  {author} {\bibinfo {author} {\bibfnamefont {E.}~\bibnamefont
  {Garrido}}\ and\ \bibinfo {author} {\bibfnamefont {A.~S.}\ \bibnamefont
  {Jensen}},\ }\href {\doibase 10.1103/PhysRevC.101.034003} {\bibfield
  {journal} {\bibinfo  {journal} {Phys. Rev. C}\ }\textbf {\bibinfo {volume}
  {101}},\ \bibinfo {pages} {034003} (\bibinfo {year} {2020})}\BibitemShut
  {NoStop}%
\bibitem [{\citenamefont {Garrido}\ \emph {et~al.}(2002)\citenamefont
  {Garrido}, \citenamefont {Fedorov},\ and\ \citenamefont
  {Jensen}}]{Garrido2002}%
  \BibitemOpen
  \bibfield  {author} {\bibinfo {author} {\bibfnamefont {E.}~\bibnamefont
  {Garrido}}, \bibinfo {author} {\bibfnamefont {D.~V.}\ \bibnamefont
  {Fedorov}}, \ and\ \bibinfo {author} {\bibfnamefont {A.~S.}\ \bibnamefont
  {Jensen}},\ }\href {\doibase 10.1016/S0375-9474(02)01020-5} {\bibfield
  {journal} {\bibinfo  {journal} {Nucl. Phys. A}\ }\textbf {\bibinfo {volume}
  {708}},\ \bibinfo {pages} {277} (\bibinfo {year} {2002})}\BibitemShut
  {NoStop}%
\bibitem [{\citenamefont {Spyrou}\ \emph {et~al.}(2010)\citenamefont {Spyrou}
  \emph {et~al.}}]{Spyrou2010}%
  \BibitemOpen
  \bibfield  {author} {\bibinfo {author} {\bibfnamefont {A.}~\bibnamefont
  {Spyrou}} \emph {et~al.},\ }\href {\doibase 10.1016/j.physletb.2009.12.016}
  {\bibfield  {journal} {\bibinfo  {journal} {Phys. Lett. B}\ }\textbf
  {\bibinfo {volume} {683}},\ \bibinfo {pages} {129} (\bibinfo {year}
  {2010})}\BibitemShut {NoStop}%
\bibitem [{\citenamefont {Gogny}\ \emph {et~al.}(1970)\citenamefont {Gogny},
  \citenamefont {Pires},\ and\ \citenamefont {Tourreil}}]{GPT}%
  \BibitemOpen
  \bibfield  {author} {\bibinfo {author} {\bibfnamefont {D.}~\bibnamefont
  {Gogny}}, \bibinfo {author} {\bibfnamefont {P.}~\bibnamefont {Pires}}, \ and\
  \bibinfo {author} {\bibfnamefont {R.~D.}\ \bibnamefont {Tourreil}},\ }\href
  {\doibase 10.1016/0370-2693(70)90552-6} {\bibfield  {journal} {\bibinfo
  {journal} {Phys. Lett. B}\ }\textbf {\bibinfo {volume} {32}},\ \bibinfo
  {pages} {591} (\bibinfo {year} {1970})}\BibitemShut {NoStop}%
\bibitem [{\citenamefont {Garrido}\ \emph
  {et~al.}(2004{\natexlab{b}})\citenamefont {Garrido}, \citenamefont
  {Fedorov},\ and\ \citenamefont {Jensen}}]{Garrido2004b}%
  \BibitemOpen
  \bibfield  {author} {\bibinfo {author} {\bibfnamefont {E.}~\bibnamefont
  {Garrido}}, \bibinfo {author} {\bibfnamefont {D.~V.}\ \bibnamefont
  {Fedorov}}, \ and\ \bibinfo {author} {\bibfnamefont {A.~S.}\ \bibnamefont
  {Jensen}},\ }\href {\doibase 10.1103/PhysRevC.69.024002} {\bibfield
  {journal} {\bibinfo  {journal} {Phys. Rev. C}\ }\textbf {\bibinfo {volume}
  {69}},\ \bibinfo {pages} {024002} (\bibinfo {year}
  {2004}{\natexlab{b}})}\BibitemShut {NoStop}%
\bibitem [{\citenamefont {Raynal}\ and\ \citenamefont {Revai}(1970)}]{RR70}%
  \BibitemOpen
  \bibfield  {author} {\bibinfo {author} {\bibfnamefont {J.}~\bibnamefont
  {Raynal}}\ and\ \bibinfo {author} {\bibfnamefont {J.}~\bibnamefont {Revai}},\
  }\href {\doibase 10.1007/BF02756127} {\bibfield  {journal} {\bibinfo
  {journal} {Nuovo Cim.}\ }\textbf {\bibinfo {volume} {68A}},\ \bibinfo {pages}
  {612} (\bibinfo {year} {1970})}\BibitemShut {NoStop}%
\bibitem [{\citenamefont {Hiyama}\ \emph {et~al.}(2019)\citenamefont {Hiyama},
  \citenamefont {Lazauskas}, \citenamefont {Marqu\'es},\ and\ \citenamefont
  {Carbonell}}]{Hiyama19}%
  \BibitemOpen
  \bibfield  {author} {\bibinfo {author} {\bibfnamefont {E.}~\bibnamefont
  {Hiyama}}, \bibinfo {author} {\bibfnamefont {R.}~\bibnamefont {Lazauskas}},
  \bibinfo {author} {\bibfnamefont {F.~M.}\ \bibnamefont {Marqu\'es}}, \ and\
  \bibinfo {author} {\bibfnamefont {J.}~\bibnamefont {Carbonell}},\ }\href
  {\doibase 10.1103/PhysRevC.100.011603} {\bibfield  {journal} {\bibinfo
  {journal} {Phys. Rev. C}\ }\textbf {\bibinfo {volume} {100}},\ \bibinfo
  {pages} {011603} (\bibinfo {year} {2019})}\BibitemShut {NoStop}%
\end{thebibliography}%

\end{document}